\documentclass[12pt]{article}
\usepackage{epsfig}
\usepackage{amsfonts,amssymb}
\newcommand{\vvec}   {\mbox{$\vec{v}$}}
\newcommand{\pip}    {\mbox{$\pi^{+}$}}
\newcommand{\pim}    {\mbox{$\pi^{-}$}}
\newcommand{\piz}    {\mbox{$\pi^{0}$}}

\newcommand{\sigbar} {\mbox{$\bar \sigma$}}

\begin{document}
\begin{sloppypar}

\begin{center}
\begin{Large}
\bf {Single} {$\mathbf{\pim}$ \bf {production in np collisions for excess 
energies up to 90 MeV}}

\end{Large}
\vspace*{5mm}

COSY-TOF collaboration

M.~Abdel-Bary$^d$, K.-Th.~Brinkmann$^a$, H.~Clement$^c$, E.~Doroshkevich$^c$,
S.~Dshemuchadse$^a$, A.~Erhardt$^c$, W.~Eyrich$^b$, H.~Freiesleben$^a$, 
A.~Gillitzer$^d$, R.~J\"akel$^a$, L.~Karsch$^a$, K.~Kilian$^d$, 
E.~Kuhlmann$^{a,\#}$, K.~M\"oller$^e$, H.P.~Morsch$^{d,f}$, L.~Naumann$^e$, 
N.~Paul$^d$, C.~Pizzolotto$^b$, J.~Ritman$^d$, E.~Roderburg$^d$, W.~Schroeder$^b$, 
M.~Schulte-Wissermann$^a$, Th.~Sefzick$^d$, A.~Teufel$^b$, A.~Ucar$^d$, 
P.~Wintz$^d$, P.~W\"ustner$^d$, and P.~Zupranski$^f$

\end{center}

%\vspace*{3mm}

\begin{footnotesize}

$^a$ Institut f\"ur Kern- und Teilchenphysik, Technische
Universit\"at Dresden, D-01062 Dresden, Germany\\
$^b$ Physikalisches Institut, Universit\"at Erlangen, D-91058
Erlangen, Germany\\
$^c$ Physikalisches Institut, Universit\"at T\"ubingen, D-72076
T\"ubingen, Germany\\
$^d$ Institut f\"ur Kernphysik, Forschungszentrum J\"ulich, 
D-52425 J\"ulich, Germany\\
$^e$ Institut f\"ur Strahlenphysik, Forschungszentrum 
Dresden-Rossendorf, D-01314 Dresden, Germany\\
$^f$ Andrzej Soltan Institute for Nuclear Studies, PL-00681, 
Warsaw, Poland

\end{footnotesize}

\begin{center}
August 15, 2007\\

\end{center}
\vspace*{5mm}

{\bf Abstract}
The quasifree reaction $np\to pp\pim$ was studied in a kinematically complete 
experiment by bombarding a liquid hydrogen target with a deuteron beam of 
momentum 1.85\,GeV/c and analyzing the data along the lines of the spectator 
model. In addition to the three charged ejectiles the spectator proton was 
also detected in the large-acceptance time-of-flight spectrometer COSY-TOF. It 
was identified by its momentum and flight direction thus yielding access to the 
Fermi motion of the bound neutron and to the effective neutron 4-momentum 
vector $\mathbb{P}_n$ which differed from event to event. A range of almost 
90\,MeV excess energy above threshold was covered. Energy dependent angular
distributions, invariant mass spectra as well as fully covered Dalitz plots 
were deduced. Sizeable $pp$ FSI effects were found as were contributions of $p$ 
and $d$ partial waves. The behavior of the elementary cross section $\sigma_{01}$ 
close to threshold is discussed in view of new cross section data. In comparison
with existing literature data the results provide a sensitive test of the 
spectator model. 

\begin{small}

\vspace{3mm}

$\#$ corresponding author, e-mail: \tt e.kuhlmann@physik.tu-dresden.de \rm\\ 
\end{small}

\newpage

\noindent

\section {Introduction}

The low-energy regime in strong interaction physics is known to be dominated by
pion producing reactions. Pioneering experiments were first carried out with
bubble chambers which somewhat later were superseded by more refined experiments
with electronic detectors at such hadron facilities as TRIUMF, LAMPF, PSI, and 
SATURNE and then, starting in the late 1980s, at the cooler rings IUCF, CELSIUS,
and COSY. A vast amount of data on meson production in general and on pion 
production in particular became available, as can be seen in the recent review 
articles from the experimental \cite{pawel} and the theoretical side 
\cite{chris}. Assuming isospin invariance, all single pion production reactions 
in $NN$ collisions with three body final states can be decomposed into a sum of 
at most two out of three elementary cross sections $\sigma_{I_{i},I_{f}}$ 
\cite{rose} with $I_i$ and $I_f$ denoting the isospin of the $NN$ system in the 
initial and final state, respectively. Starting from the $pp$-entrance channel 
two reactions are possible, $pp \to pp\piz$ and $pp \to pn\pip$, yielding,
respectively, $\sigma_{11}$ and $\sigma_{11}+\sigma_{10}$. The isoscalar cross 
section $\sigma_{01}$ can only be determined in neutron-proton collisions as, 
$e.g.$, $\sigma_{np\to pp\pi^{-}}\,=\,\frac{1}{2}(\sigma_{11}+\sigma_{01})$ and 
hence

\vspace{-8mm}
\begin{center}
\begin{equation}
\sigma_{01} = 2\cdot \sigma_{np\to pp\pi^-} - \sigma_{pp \to pp\pi^0}.
\end{equation}
\end{center}

Close to threshold only a few partial waves have to be considered. In a simple,
semiclassical picture the maximum $l\,=\,l_{max}$ is given by $l_{max} \simeq
R\cdot q$. Here $R$, which gives the distance from the center of the $NN$ 
collision where a pion of momentum $q$ is created, is of order $h/(m_{\pi}c)$,
$i.e.$, the compton wavelength of the pion, and hence $l_{max}\simeq q/m_{\pi}
\simeq \eta$. This dimensionless parameter $\eta$ is often found in the 
literature for specifying excess above threshold. A more direct measure 
is the excess energy $Q\,=\,\sqrt{s}-\sqrt{s_0}$ with $\sqrt{s} (\sqrt{s_0})$ 
denoting the total CM energy (the threshold energy), respectively. Since for 
excess energies less than 100 MeV $\eta$ is well below 1.5, only partial waves 
of type $Ss, Sp, Ps$ and $Pp$ as, $e.g.$, $^3P_0\to$$ ^1S_0s_0$ have to be 
considered, whereas the role of $l\,=\,2$ contributions ($Sd$ and $Ds$) can be 
regarded as negligible. Here the Rosenfeld notation $L_{p}l_{q}$ has been used 
\cite{rose} with $L_p$ being the orbital angular momentum of the $NN$ pair, 
$l_q$ that of the pion with respect to this pair.

In the past the $pp$ entrance channel attracted most of the attention. Only
recently the focus was shifted to a larger degree towards the study of
proton-neutron collisions. In this case one might either use a neutron beam 
impinging on a hydrogen target or use a deuterium target as a substitute for a 
neutron and use the spectator model to determine the observables of the quasifree
reaction. The former approach was extensively used by the Freiburg group working 
at PSI \mbox{$\lbrack$4-7$\rbrack$} and to some extent by groups at LAMPF 
\cite{tomas} and TRIUMF \cite{bach}. The first dedicated experiments on a 
deuterium target were also performed at TRIUMF \mbox{$\lbrack$10,11$\rbrack$} 
where in order to bypass the difficulties of low-energy spectator detection, the 
final-state protons were restricted to the $^1S_0$, the ``diproton'' state, by 
selecting only two-proton events with a small relative momentum. Alternatively 
one can use a deuteron beam hitting a hydrogen target. The advantage of 
having fast spectator protons flying in the forward direction is counterbalanced
by the fact that appropriate deuteron beams beyond 4\,GeV/c are not available 
any more and thus one is restricted to reactions where only light mesons are 
produced. The feasibility of this approach was investigated in our recent paper 
\cite{spect} on the $dp\to ppp\pim$ reaction.

It is the aim of the present paper on the quasifree $np\to pp\pim$ reaction to 
present angular distributions, Dalitz plots and invariant mass distributions 
for the three reaction products, from which one might deduce clues as to the 
participating partial waves.  We will present our data for angles as defined in 
Fig.~1. In the 3-particle center-of-mass (CM) system all three momentum vectors 
lie in a plane, their sum adds up to zero. Observables when given in the CM 
system will be marked with an asterisk. Denoting the beam direction by $\vec{N}$,
the pion momentum by $\vec{q}^*$ and the two proton momenta by $\vec{p_i^*}$, 
we will present angular distributions in the two angles cos $\theta_q^*$\,=\,cos
($\vec{q^*},\vec{N}$) and cos $\theta_P^*$\,=\,cos ($\vec{P^*},\vec{N}$) where 
$\vec{P^*}$, the relative proton momentum, is taken as the difference vector 
$\vec{p_2^*}-\vec{p_1^*}$. By construction the latter is symmetric around 
cos $\theta_P^*$\,=\,0. Strong interference effects between outgoing $Ss$ and 
$Sp$ on one hand and $Ps$ and $Pp$ waves on the other will produce highly 
asymmetric distributions in cos $\theta_q^*$ even at the lowest $Q (<$20\,MeV).

\vspace*{-4.0cm}
\begin{figure}[htb]
\begin{center}
\epsfig{file=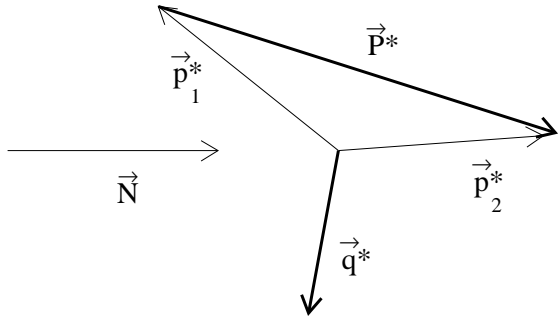,scale=0.75}
\end{center}
\vspace*{-6.0cm}
\caption{\it Definition of the relevant momentum vectors in the CM system.}
\end{figure}

\section{Experimental Procedure}

\subsection{Principle of measurement and spectator tagging}

The reaction under study is $np\to pp\pim$. Due to the lack of a neutron beam, 
deuterons were used instead and the data were analyzed along the lines of the 
spectator model. This method has been described in detail in 
\cite{spect}, hence we will give only some short remarks here. The basic idea 
of the model is that 1) the proton in the deuteron can be regarded as an 
unaffected spectator staying on-shell throughout the reaction and 2) the matrix 
element for quasi-free pion production from a bound neutron is identical to that
for free pion production from an unbound neutron. Crucial to the method is the 
task to detect and identify the spectator proton $p_s$, since the information 
gathered from this particle gives a direct measure of the Fermi momentum carried 
by the off-shell neutron within the deuteron at the time of the $np$ reaction.
The Fermi momentum distribution as calculated from any of the existing $NN$ 
potentials has a maximum near 40~MeV/c and a tail extending towards several 
hundred MeV/c, hence a wide range in excess energy $Q$ can be covered with a 
monoenergetic deuteron beam. The main result of our recent study was that the 
two assumptions quoted above can be regarded as being fulfilled for Fermi 
momenta below 150\,MeV/c.

The experiment was carried out with the time-of-flight spectrometer COSY-TOF set 
up on an external beamline of the proton synchrotron COSY \cite{rudi} at the 
Forschungszentrum J\"ulich. A deuteron beam of momentum $p_d$\,=\,1.85\,GeV/c
was focussed onto a liquid hydrogen target, charged particles emerging from the 
reaction zone were detected in a multi-layer scintillator hodoscope with its 
main components Quirl, Ring and Barrel. Details of the various subdetectors, 
their performance as well as the different steps necessary for calibrating the 
whole system have been described in a series of papers, see \cite{spect} 
and references therein. Here only a short overview will be given. By measuring 
each particle's flight time and direction, their velocity vectors given as 
$\vvec = (\beta, \theta, \phi)$ could be determined with a time-of-flight 
resolution of better than 300\,ps $(\sigma )$ and an angular track resolution of 
better than 0.3$^\circ (\sigma )$. The momentum 4-vectors $\mathbb{P}$ of all 
detected particles were then obtained from the measured observables by applying 
additional mass hypotheses. Carrying out various tests as, $e.g.$, momentum 
conservation, missing mass and invariant mass analyses as well as comparisons 
with results obtained from our Monte Carlo simulations helped to find the correct
assignment for each event with a high degree of probability as quantified below.

In the reaction $dp\to pp\pim p_s$ four charged particles are emitted which in
most cases all are detected in the time-of-flight spectrometer. Thus the main
trigger condition was such that a total of four hits was required in any of the
stop scintillator hodoscopes Quirl, Ring and Barrel and at least one hit in
the twelve-fold segmented start scintillator. Due to the fact that pions can
also be emitted into the backward region where no detector was installed we set
up a second trigger condition with only three required hits at a reduction 
factor of 10. Since for these events the unobserved pion can be reconstructed
through a missing mass analysis, the full kinematically allowed phase space was 
covered. 

With a beam momentum below the threshold for $2\pi$-production, any 4-hit 
event apart from accidentals could only result from the reaction under study. 
As the first step in our analysis we checked on the four possible hypotheses, 
$i.e.$, the pion being particle 1, 2, 3, or 4 and calculated for each case the 
sums of longitudinal and transversal momentum components $\sum p_L$ and 
$\sum p_T$. As the correct assignment we took the one where these values were 
closest to $\sum p_L$\,=\,p$_d$ and $\sum p_T$\,=\,0. As the spectator proton 
we then chose the one which was detected close to the beam axis with a 
momentum near p$_d$/2. The spread in Fermi momentum caused the momentum of the 
spectator to vary considerably, higher $Q$ values correspond to lower 
momenta and vice versa. This is illustrated in Fig.~2 where for two narrow 
ranges in $Q$ ($\lbrack$18.0-34.0$\rbrack$ and $\lbrack$61.0-74.5$\rbrack$\,MeV) 
the momentum distribution of the spectator proton and the summed distribution of 
both reaction protons is plotted. The spectator distribution given by the solid 
histogram at a mean $<Q>$\,=\,26 MeV sticks out as a sharp line well separated 
from the much broader momentum distribution of the other two protons, whereas at 
$<Q>$\,=\,68 MeV the spectator line is still rather narrow, but starts to 
overlap with the one of the reaction protons. Since a unique identification of 
the spectator is essential for the analysis we found it necessary to also limit 
the range in $Q$ due to this effect and only considered events where the excess 
energy was below 90\,MeV which roughly coincides with our proposed limit for 
the Fermi momentum \cite{spect}. In our finally accepted data sample of 
2.2$\cdot$10$^5$ events we obtained a longitudinal momentum distribution 
$\sum p_L$ which had its center at $p_d$ with a width of 39\,MeV/c ($\sigma$). In
case of the transversal distribution the spread was even smaller, namely 
13\,MeV/c. Alternatively we also used the missing mass method for identifying
the various ejectiles, see ref.\cite{spect}, and found full agreement.

\begin{figure}[htb]
\vspace*{-1.0cm}

\hspace{5cm}
%\begin{center}
\epsfig{file=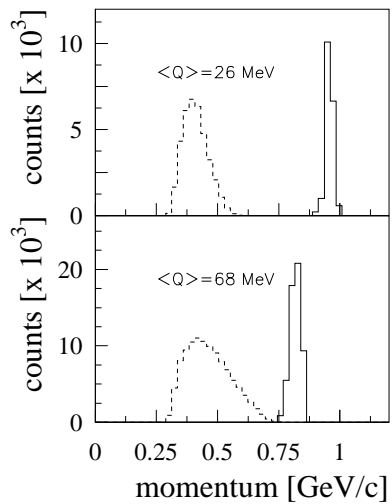,scale=0.55}
%\end{center}
\vspace*{-3.5cm}
\caption{\it Experimentally deduced momentum distributions for the spectator 
protons (solid histograms) and the reaction protons (dashed histograms) for two 
ranges in $Q$.}

\end{figure}

Reconstructing those events where the pions were emitted into the backward region
by missing mass techniques in principle caused no problems. In several cases, 
however, we found events where three protons were detected as for a true $dp\to 
pp\pim p_s$ reaction, only there a third proton was produced in a chain of two 
consecutive $np$ elastic scattering processes. From the first quasielastic 
scattering reaction one gets a scattered proton, a forward flying spectator and 
a neutron. The scattered neutron in traversing one of the start detector 
elements hits another proton which reaches the detector whereas the slowed-down 
neutron remains unobserved. In simulating this process we found that by suitable
selections in missing mass and angles these events could be eliminated. Thus an
additional set of roughly 0.6$\cdot 10^5$ reconstructed $pp\pim p_s$ events was
obtained.

As has been outlined in \cite{spect}, the timing signals deduced from both ends 
of the Barrel scintillators not only yield information on the flight times, but 
also on the hit position of any track passing through the Barrel. Hence an 
important step in the detector calibration is the fixing of the absolute time 
offset which was carried out through a comparison with the results obtained for 
$dp$ elastic scattering. This binary reaction with its unique kinematics and 
sizeable cross section was repeatedly measured in separate runs with an adjusted 
trigger condition. As a check of the reliability of the event reconstruction we 
show in Fig.~3 the deuteron angular distribution (given as histograms) in 
comparison with older data obtained at a somewhat higher beam momentum (solid 
dots) \cite{boot}. Instead of the deuteron beam momentum of 1.85\,GeV/c we quote 
a value of 0.92\,GeV/c ($T_{kin}$\,=\,376\,MeV) which corresponds to the 
inverse reaction for a proton beam hitting a deuterium target at the same 
$\sqrt{s}$. The absence of data in the forward region is due to the fact that 
the corresponding protons were emitted towards angles $>60^\circ$ which is out 
of the acceptance of the spectrometer. The overall agreement is very good, the 
apparent mismatch in the peaking of the forward maximum results from the 
difference in beam momentum $p_p$. To eliminate this dependency on $p_p$ we

\begin{figure}[htb]
\vspace*{-1.0cm}
\begin{center}
\epsfig{file=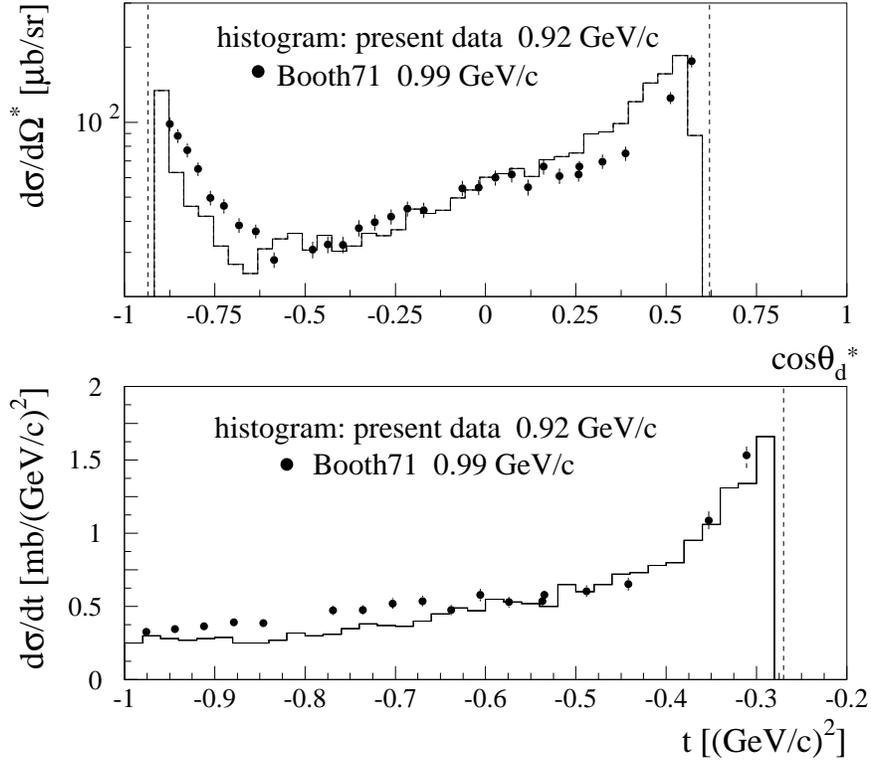,scale=0.6}
\end{center}
\vspace*{-1.4cm}
\caption{\it Angular distributions for the elastic $pd$ scattering reaction 
plotted as a function of the CM angle cos $\theta^{*}_d$ (top) and momentum
transfer t (bottom) in comparison with data from ref.\cite{boot}. The dashed 
vertical lines denote the acceptance limits of our detector.}
\end{figure}

plotted the same data (for cos $\theta^*_d > -0.35$) as a function of the 
Mandelstam variable t (Fig.~3, lower frame) and found a very satisfying agreement. 

\subsection{Monte Carlo simulation}

The analysis of our experimental data samples was accompanied by extensive Monte 
Carlo simulations. In order to allow each simulated quasi-free $np\to 
pp\pim$-event to have different initial kinematical parameters the program 
package was modified in a way as was described in detail in \cite{spect}, hence 
we will give only a short outline of the main ideas. Using the CERNLIB event 
generator GENBOD \cite{gb} one generates $N$-body events for a given reaction 
specified by $N$, type and mass of the particles involved and the total CM energy
$\sqrt{s}$. The code returns momentum 4-vectors for each ejectile in the overall 
center-of-mass system and weight factors $w_e$ based on the phase space density 
of the reaction. In the present case the basic reaction to be simulated is $np 
\to pp\pim$. For each event randomly chosen values for cos $\theta^{*}$, 
$\phi^{*}$ and momentum $|\vec{p}^{*}|$/(MeV$\cdot$c$^{-1})$ were picked, the 
two former ones following uniform distributions, whereas the momentum was folded 
with the above mentioned Fermi distribution. We identify the three-component 
vector {$|\vec{p}^{*}|$, cos $\theta^{*}, \phi^{*}$} as well as the one pointing 
into the opposite direction with those of an $np$-pair within the deuteron in 
its CM system. Transformation into the laboratory system then allows one to 
deduce the corresponding vectors for spectator and projectile particle within 
a fast moving deuteron of momentum p$_d$\,=\,1.85\,GeV/c. The fact that in the 
laboratory system the flight direction of the projectile neutron deviates 
by a small angle from that of the beam deuteron is accounted for by a 
suitably chosen rotation such that the neutron's flight direction serves 
as the actual beam direction. After having fixed event-by-event the 
momentum vector for the ``beam neutron'' it is straightforward to perform 
the simulation for $np \to pp \pim$.

By using approximately 1 million Monte Carlo events uniformly distributed across 
the available phase-space we could determine the energy dependent acceptance of 
our detector and the reconstruction efficiency as a function of excess energy 
$Q$. The main limitations in acceptance stemmed from the maximum in detector 
angle $\theta_{max}\,=\,60^{\circ}$ and from the charged particles' energy loss 
in the various detector layers resulting in a low $\beta$-threshold of $\beta 
\approx$\,0.5 for $\pim$ mesons and $\beta \approx$\,0.35 for protons. In Fig.~4 
we show the resulting acceptance curves for the relative proton momentum angle 
cos $\theta_P^*$ (left) and the proton-proton invariant mass 
 
\begin{figure}[!htb]
\vspace*{-0.6cm}
\begin{center}
\epsfig{file=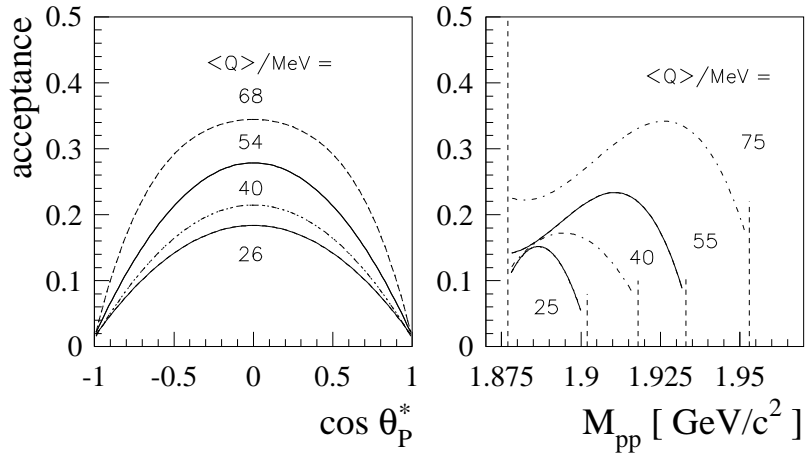,scale=0.65}
\end{center}
\vspace*{-6.5cm}
\caption{\it Simulated acceptance curves as obtained for the relative proton 
momentum angle \mbox{cos $\theta^*_P$} (left) and the proton-proton invariant 
mass $M_{pp}$ (right) for different values of excess energy $Q$. The dashed 
vertical lines in the plot of the right panel denote the kinematical limits.}
\end{figure}

\newpage 

$M_{pp}$ (right) for selected values of $Q$. The acceptance goes to zero 
near $|$cos $\theta_P^*|\,=\,1$. This comes as a result of the way the relative 
proton momentum vector $\vec{P^*}$ is constructed (see also Fig.~1). When the 
direction of $\vec{P^*}$ approaches the beam direction, one of the protons in 
the CM system moves backwards, hence has minimum energy and drops below the 
detection threshold. Similar calculations have been performed for all other 
observables.

\section{Results and discussion}

Cross sections for various emission angles and invariant masses of each 
two-particle subsystem as well as two-dimensional Dalitz plots were extracted 
from the data. In order to derive absolute cross sections one must know the 
integrated luminosity. Defined as $\cal{L}$ =$ \int n_b \cdot n_t\,dt$ with 
$n_b (n_t)$ denoting the number of beam and target particles, respectively, one 
finds the cross section from the relation 

\vspace{-8mm}
\begin{center}
\begin{equation}
\sigma = n/(\mathcal{L}\cdot f \cdot \epsilon),
\end{equation}
\end{center}

where $n$ is the number of observed events, $f$ the deadtime correction factor 
and $\epsilon$ gives the geometrical and reconstruction efficiency. This simple 
relation, however, has to be modified in case of a quasifree reaction. The Fermi 
motion of the neutrons within the deuteron will lead to a wide span in excess 
energy $Q$ such that the number of beam particles initiating a $pn$ reaction at 
a given $Q$ will vary. In the present case of a close-to-threshold measurement, 
one furthermore will observe a strong variation in $\sigma$. In order to extract 
the energy dependence of the cross section and to compare it with the one of the 
free reaction it is necessary to unfold the effect of the Fermi motion from the 
data. By dividing the range in $Q$ in small bins $<Q>_i$ such that per bin the 
variation in $\sigma$ is small and can be approximated by a constant $\sigbar_i$ 
the number of produced events $N_i$ is \cite{stina}

\vspace{-8mm}
\begin{center}
\begin{equation}
N_i = \sigbar_i \mathcal{L}\int_{<Q>_i} |\phi(p_b)| d^3\mathbf{p}_b,
\end{equation}
\end{center}

where the integral is taken over all neutron beam momenta $\mathbf{p}_b$ 
contributing to $<Q>_i$ and $\phi(p_b)$ is the deuteron wave function as
given by the PARIS potential \cite{paris}. Here $\cal{L}$ is again the 
overall luminosity, its $Q$-dependence is accounted for by the integral. 
Correspondingly the number of observed events is given by

\vspace{-8mm}
\begin{center}
\begin{equation}
n_i = N_i\cdot f \cdot \epsilon_i
\end{equation}
\end{center}

The evaluation of the integral is performed by means of Monte Carlo simulations. 
Denoting the total number of generated Monte Carlo events by $N^{MC}$ and the one
generated for the bin $<Q>_i$ by  $N_i^{MC}$ the integral is given by the ratio

\vspace{-8mm}
\begin{center}
\begin{equation}
\int_{<Q>_i} |\phi(p_b)| d^3\mathbf{p}_b = \frac{N_i^{MC}}{N^{MC}}.
\end{equation}
\end{center}

%\newpage

Finally, by using eqs.~$\lbrack$3-5$\rbrack$ one finds for the cross section

\vspace{-8mm}
\begin{center}
\begin{equation}
\bar{\sigma _i} = \frac{1}{\mathcal{L}\cdot f \cdot \epsilon_i} \cdot 
\frac{n_i}{N_i^{MC}/N^{MC}}.
\end{equation}
\end{center}

Defining $\tilde n_i$\,=\,$n_i/\epsilon_i$ as the number of observed and acceptance
corrected events the cross section is essentially given as $\bar{\sigma _i} 
\propto \tilde n_i/N_i^{MC}$ since $\cal{L}$, $f$ and $N^{MC}$ are constants. 
This is demonstrated in Fig.~5 where we show the distribution of observed events 
in the top frame as the solid histogram; it extends from threshold up to 90\,MeV.
Also shown in this frame is the corresponding distribution as obtained for our 
Monte Carlo events. Calculated for a deuteron beam momentum of 
$p_d$\,=\,1.85\,GeV/c it has its largest $Q$ values also near 90\,MeV, but on 
the low side starts with a sizeable yield already at threshold. Its maximum is 
shifted to lower $Q$ values towards the peak of the deuteron wavefunction. 
When extracting the ratio of the experimentally deduced distribution and the 
Monte Carlo data, the histogram as shown in the bottom frame is obtained which 
(note the logarithmic scale) rises by more than two orders of magnitude. Also
shown as a dashed curve are the total cross section data obtained with a free 
neutron beam at PSI \cite{daum1} and parameterized as a $3^{rd}$ order polynomial
in $Q$. When applying a suitably chosen normalization factor the present data 
are in good agreement with these absolute values with only some minor deviations 
at the lower and upper ends of the covered range. Henceforth this one 
normalization factor will be used in all of our further presentations and 
discussions of differential cross sections.

\begin{figure}[htb]
\vspace*{-1.2cm}

\hspace{5cm}
%\begin{center}
\epsfig{file=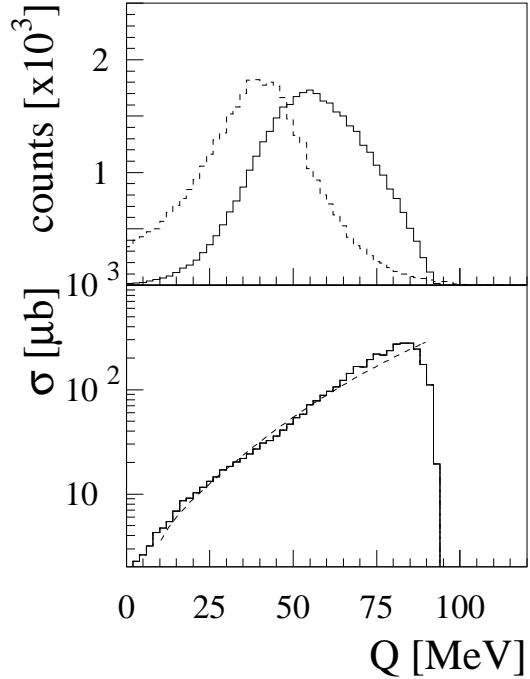,scale=0.75}
%\end{center}
\vspace*{-5.0cm}
\caption{\it Top: Simulated (dashed, MC) and measured (solid histogram, exp) 
distributions of $pp\pim$ events as a function of Q. Bottom: Ratio of the two 
distributions exp/MC (solid histogram) normalized to the cross section data 
measured at PSI \cite{daum1} (dashed curve).} 

\end{figure}

No attempt was made to derive in an independent way absolute cross sections from 
the present experiment. The natural choice for the determination of the 
luminosity $\cal{L}$ would have been the $pd$ elastic scattering reaction. 
Although it was quite successfully used for calibration procedures we did not 
consider it as being suited for finding the size of $\cal{L}$. Firstly the 
amount of available cross section data is still scarce. Apart from the already 
mentioned experiment by Booth et al.~\cite{boot} at $p_p$\,=\,0.99\,GeV/c which
corresponds to a bombarding energy of 425\,MeV we only found one more set of 
published data by Alder et al.~\cite{ald} at comparable energies. These authors 
present data at proton bombarding energies of 316 and 364 MeV covering far 
backward angles cos $\theta^* <-0.6$ and at 470 and 590 MeV at angles 
cos $\theta^* <0$. Due to this small range in cos $\theta^*$ we consider these 
data unfit for a reliable interpolation. Recent data by G\"ulmez et al.
\cite{gul} were taken at much higher energies of 641 and 793\,MeV, those by 
Rohdje\ss{} et al.~\cite{rohd} at energies up to 300\,MeV. Secondly we found it 
difficult to estimate the error in $\sigma_{pd}$ when extracting the $dp$ 
elastic events from the underlying background which was dominated by the much 
stronger $pp$ quasielastic scattering events. Finally the uncertainties due to 
effects like shadowing and rescattering which tend to reduce the cross sections 
of any quasifree reaction by about $8\%$ \cite{chia} and thus add to the size of 
the systematic error, would only then be of minor consequence when a comparison 
with another quasifree reaction is carried out.

An order of magnitude estimate of the cross section could nevertheless be made.
The integrated luminosity was found from the known target thickness (4~mm
liquid hydrogen corresponding to $1.8\cdot 10^{22}$/cm$^2$ target particles), the 
average beam intensity of $7\cdot 10^6$ deuterons/s, and the total running time
to be of order 40\,nb$^{-1}$. Using eq.~2 with $f\approx 0.5$ and $\epsilon 
\approx 0.25$ one finds a mean cross section near 60\,$\mu$b in good agreement 
with the PSI data \cite{daum1}.
 
\subsection{Angular distributions}

Acceptance corrected angular distributions are shown in Figs.~6 and 7 together
with fits in terms of Legendre polynomials. The excess energy range 1.0-88.0
MeV was cut into six bins, namely $\lbrack$1.0-18.0$\rbrack$, $\lbrack$18.0-34.0
$\rbrack$, $\lbrack$34.0-47.5$\rbrack$, $\lbrack$47.5.0-61.0$\rbrack$, $\lbrack$
61.0-74.5$\rbrack$, and $\lbrack$74.5.0-88.0$\rbrack$ MeV, the indicated excess 
energies $<Q>$ denote the center values of these bins. Error bars when given 
denote statistical errors only. It should be kept in mind that, although the 
cross 

\begin{figure}[!htb]
\vspace*{-0.5cm}
\begin{center}
\epsfig{file=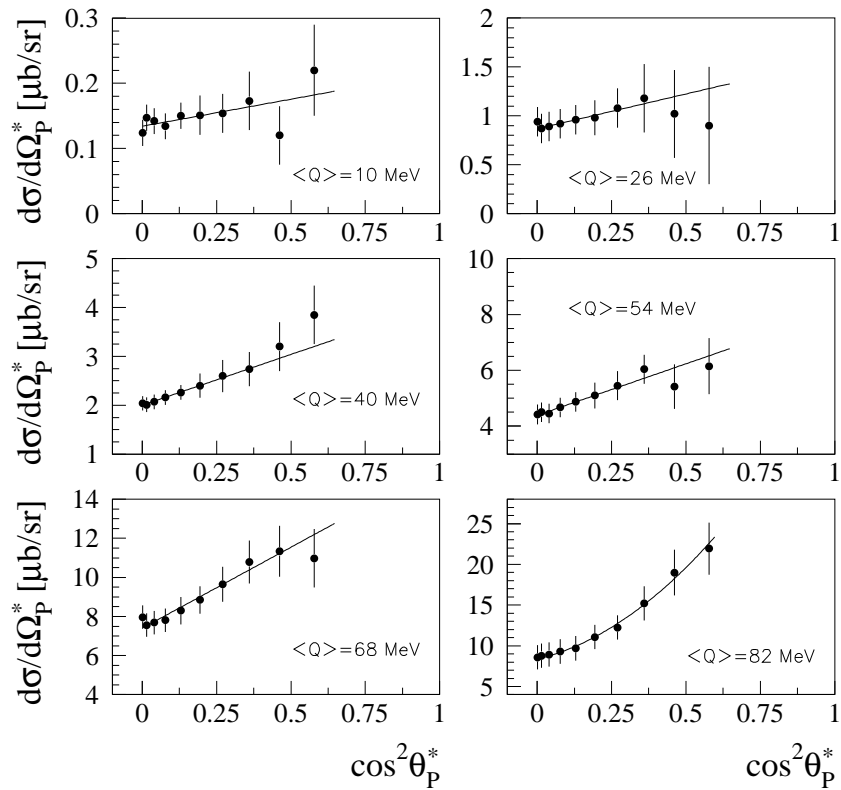,scale=0.60}
\end{center}
\vspace*{-0.8cm}
\caption{\it Angular distributions of the relative proton momentum for selected 
excess energies $<$Q$>$ together with results of Legendre polynomial fits.}  
\end{figure}

section rises monotonically with $Q$, the observed counting rate does 
not. Due to the non-uniform Fermi distribution which governs the available excess 
energies, the highest rates are found near $Q$\,=\,50\,MeV and consequently one 
also observes the lowest statistical errors there. As already outlined above 
(see Fig.~1) the angular distributions of the relative proton momentum by 
construction are symmetric with respect to cos $\theta_P^*$\,=\,0. They are 
plotted as a function of cos$^2\theta_P^*$ and were fitted with even Legendre 
polynomials $W(cos~\theta) \propto 1 + \sum_{\nu}^{}a_{2\nu}\cdot 
P_{2\nu}(cos~\theta)$ up to $\nu\,=\,2$. The extracted expansion coefficients 
are given in Table~1, up to $<Q>$\,=\,68\,MeV the $P_4$-term was neglected. 
In addition we give the numerical values for $d\sigma /d\Omega^*$ in Table~2 
where, as mentioned before, the absolute scale was adjusted to the PSI data 
\cite{daum1}.

\begin{table}[!htb]
\begin{center}
\caption {\it Expansion coefficients of the Legendre polynomial fits to the 
angular distributions of the relative proton momentum.} 

\vspace{0,5cm}

\begin{tabular}{ccc}

\hline
%& &   \\
$<Q>/MeV$ &$a_2$ & $a_4$ \\
%& &  \\
\hline
%& &  \\
10 & 0.34$\pm$0.08 & - \\
26 & 0.43$\pm$0.07 & - \\
40 & 0.51$\pm$0.06 & - \\
54 & 0.44$\pm$0.05 & - \\
68 & 0.54$\pm$0.05 & - \\
82 & 1.28$\pm$0.13 & 0.37$\pm$0.13 \\
%& &  \\
\hline

\end{tabular}
\end{center}
\end{table}

\vspace{-0.8cm}

\begin{table}[!htb]
\begin{center}
\caption {\it Differential cross sections in $\mu b/sr$ for the angular 
distributions of the relative proton momentum at six excess energies Q.}

\vspace{0,5cm}

\begin{tabular}{ccccccc}

\hline
cos$^2\theta^*_P$ & 10~MeV & 26~MeV & 40~MeV & 54 MeV & 68~MeV & 82~MeV \\
\hline
0.002 & 0.124$\pm$0.022 & 0.94$\pm$0.14 & 2.04$\pm$0.16 & 4.42$\pm$0.34 & 7.96$\pm$0.61 & 8.58$\pm$1.31 \\
0.014 & 0.147$\pm$0.023 & 0.87$\pm$0.13 & 2.01$\pm$0.15 & 4.50$\pm$0.36 & 7.56$\pm$0.62 & 8.78$\pm$1.43 \\
0.040 & 0.142$\pm$0.024 & 0.89$\pm$0.15 & 2.07$\pm$0.16 & 4.45$\pm$0.36 & 7.69$\pm$0.64 & 8.93$\pm$1.52 \\
0.078 & 0.134$\pm$0.024 & 0.92$\pm$0.16 & 2.16$\pm$0.17 & 4.67$\pm$0.38 & 7.81$\pm$0.65 & 9.29$\pm$1.48 \\
0.130 & 0.150$\pm$0.024 & 0.96$\pm$0.16 & 2.26$\pm$0.20 & 4.87$\pm$0.38 & 8.30$\pm$0.71 & 9.70$\pm$1.53 \\
0.194 & 0.151$\pm$0.028 & 0.98$\pm$0.16 & 2.40$\pm$0.25 & 5.10$\pm$0.46 & 8.85$\pm$0.73 &11.07$\pm$1.57 \\ 
0.270 & 0.154$\pm$0.033 & 1.08$\pm$0.21 & 2.60$\pm$0.33 & 5.45$\pm$0.52 & 9.65$\pm$0.92 &12.24$\pm$1.60 \\
0.360 & 0.170$\pm$0.044 & 1.18$\pm$0.34 & 2.74$\pm$0.38 & 6.04$\pm$0.54 &10.78$\pm$1.13 &15.21$\pm$2.13 \\
0.462 & 0.120$\pm$0.047 & 1.02$\pm$0.46 & 3.20$\pm$0.52 & 5.42$\pm$0.80 &11.33$\pm$1.31 &18.98$\pm$2.76 \\
0.578 & 0.222$\pm$0.071 & 0.90$\pm$0.61 & 3.85$\pm$0.63 & 6.15$\pm$1.11 &10.98$\pm$1.53 &21.94$\pm$3.12 \\ 
\hline

\end{tabular}
\end{center}
\end{table}

As one can see from inspection of Fig.~6 the scatter of the data points as well
as the size of the error bars increases drastically for values cos$^2\theta_P^* 
\geq 0.4$. This is the result of the very low acceptance observed in this angular
region (see also the discussion in context with Fig.~4). Accordingly the 
$\chi^2$ minimisation was only performed on the first eight data points. In the 
literature we found one measurement of these proton distributions, which was 
extracted from roughly 4000 bubble chamber frames \cite{hand}. At an average 
value of $Q$ near 54\,MeV the authors report a value of $a_2\,=\,0.276\pm 0.032$ 
which is to be compared to the present one of $a_2\,=\,0.44\pm 0.05$. We believe 
the observed 4$\sigma$ deviation to be due to systematic errors in their method
which were not included in the quoted error. As mentioned above only $Ss, Sp, Ps$ 
and $Pp$ partial waves should be present in the energy region covered in the 
present experiment. From our data, however, it can be seen that in the higher $Q$ 
range contributions of the $Ds$ wave are present as well. Near $<Q>$\,=\,82\,MeV 
a $P_4$(cos $\theta$) term with a sizeable expansion coefficient 
$a_4\,=\,0.37\pm0.13$ had to be included in the fit.

The pion angular distributions as deduced for the same intervals in $Q$ are 
shown in Fig.~7. In general all are asymmetric and were fitted with sizeable 
$a_1$ and $a_2$ coefficients (see Table~3). The one obtained at $<Q>$\,=\,54
MeV is compared with data taken from ref.~\cite{hand} (dotted line) and 
ref.~\cite{daum1} (dashed line) and good agreement is observed. The cross 
section as given in \cite{hand} exceeds the one measured at PSI by the factor 
1.29. The authors of ref.~\cite{daum1} explain this discrepancy with a possible 
underestimation of the mean neutron energy in the older experiment. That
measurement had been carried out over a broad neutron energy range and the 
mean energy  had been deduced from a maximum likelihood fit. In the present 
comparison the data of \cite{hand} have been rescaled to match the PSI cross 
section. For the sake of completeness we additionally give in Table 4 the 
numerical values of the differential $\pim$ cross sections. In passing we like 
to add that the corresponding ones given in ref.~\cite{daum1} (Table 3 and 
Fig.~11) are not consistent with the absolute cross section data presented in 
their Table 4, but are too low by the factor $2\pi/10$ due to an error in 
binning the data \cite{heiko}.

\begin{figure}[!htb]
\vspace*{-1.2cm}
\begin{center}
\epsfig{file=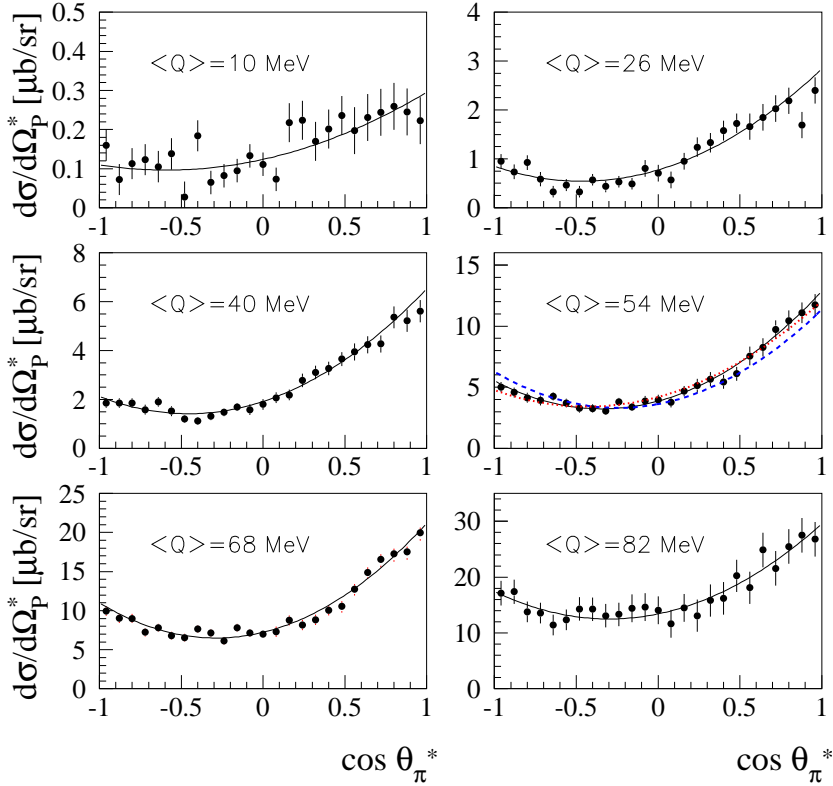,scale=0.6}
\end{center}
\vspace*{-0.8cm}
\caption{\it (Color online) Angular distributions of the pions for selected 
excess energies $<$Q$>$ together with results of Legendre polynomial fits. At 
$<$Q$>$\,=\,54 MeV, the pion angular distributions as found by Daum et al.
 \cite{daum1} and Handler \cite{hand} are also shown as a dashed (blue) and a 
dotted (red) curve, respectively.}

\end{figure}

\begin{table}[!htb]
\begin{center}

\caption {\it Expansion coefficients of the Legendre polynomial fits to the 
$\pi$ angular distributions.}

\vspace{0,5cm}

\begin{tabular}{ccc}

\hline
$<Q>/MeV$ &$a_1$ & $a_2$ \\
\hline
10 & 0.62$\pm$0.21 & 0.35$\pm$0.14  \\
26 & 0.88$\pm$0.21 & 0.63$\pm$0.28  \\
40 & 0.81$\pm$0.11 & 0.59$\pm$0.14  \\
54 & 0.65$\pm$0.07 & 0.63$\pm$0.07  \\
68 & 0.50$\pm$0.08 & 0.58$\pm$0.06  \\
82 & 0.36$\pm$0.07 & 0.40$\pm$0.06  \\
\hline

\end{tabular}
\end{center}
\end{table}

\begin{table}[!htb]
\begin{center}

\caption {\it Differential cross sections in $\mu b/sr$ of the pion angular 
distributions at six excess energies Q.}

\vspace{0,5cm}

\begin{tabular}{ccccccc}

\hline
cos $\theta^*_{\pi}$ & 10~MeV & 26~MeV & 40~MeV & 54~MeV & 68~MeV & 82~MeV \\
\hline
-0.96 &	0.160$\pm$0.041 & 0.95$\pm$0.15 & 1.85$\pm$0.21 & 5.01$\pm$0.45  &  9.95$\pm$0.61 & 17.1$\pm$2.3 \\
-0.88 &	0.072$\pm$0.039	& 0.73$\pm$0.13 & 1.85$\pm$0.20 & 4.57$\pm$0.42  &  9.01$\pm$0.60 & 17.4$\pm$2.2 \\
-0.80 &	0.113$\pm$0.041	& 0.92$\pm$0.14 & 1.85$\pm$0.20 & 4.14$\pm$0.42  &  9.00$\pm$0.58 & 13.8$\pm$1.9 \\
-0.72 &	0.123$\pm$0.038	& 0.58$\pm$0.13 & 1.58$\pm$0.18 & 3.92$\pm$0.41  &  7.23$\pm$0.55 & 13.6$\pm$1.8 \\
-0.64 &	0.105$\pm$0.042	& 0.32$\pm$0.12 & 1.90$\pm$0.19 & 4.24$\pm$0.37  &  7.82$\pm$0.54 & 11.4$\pm$1.5 \\
-0.56 &	0.138$\pm$0.037	& 0.46$\pm$0.11 & 1.52$\pm$0.18 & 3.71$\pm$0.37  &  6.80$\pm$0.52 & 12.4$\pm$1.5 \\
-0.48 &	0.027$\pm$0.038	& 0.32$\pm$0.11 & 1.20$\pm$0.16 & 3.27$\pm$0.35  &  6.55$\pm$0.53 & 14.3$\pm$1.8 \\
-0.40 &	0.184$\pm$0.036	& 0.57$\pm$0.11 & 1.12$\pm$0.16 & 3.26$\pm$0.34  &  7.65$\pm$0.52 & 14.3$\pm$1.7 \\
-0.32 &	0.065$\pm$0.034	& 0.43$\pm$0.12 & 1.31$\pm$0.15 & 3.05$\pm$0.33  &  7.14$\pm$0.50 & 13.1$\pm$1.9 \\
-0.24 &	0.082$\pm$0.031	& 0.53$\pm$0.13 & 1.47$\pm$0.14 & 3.81$\pm$0.35  &  6.12$\pm$0.50 & 13.3$\pm$2.0 \\
-0.16 &	0.095$\pm$0.032	& 0.49$\pm$0.13 & 1.69$\pm$0.16 & 3.37$\pm$0.36  &  7.82$\pm$0.52 & 14.4$\pm$2.4 \\
-0.08 &	0.133$\pm$0.029	& 0.80$\pm$0.15 & 1.58$\pm$0.17 & 3.86$\pm$0.38  &  7.14$\pm$0.53 & 14.6$\pm$2.4 \\
 0.0  &	0.111$\pm$0.028	& 0.70$\pm$0.17 & 1.80$\pm$0.19 & 3.97$\pm$0.40  &  6.97$\pm$0.52 & 14.0$\pm$2.3 \\
 0.08 &	0.073$\pm$0.033	& 0.57$\pm$0.16 & 2.07$\pm$0.21 & 3.76$\pm$0.39  &  7.31$\pm$0.58 & 11.7$\pm$2.2 \\
 0.16 &	0.218$\pm$0.038	& 0.95$\pm$0.18 & 2.18$\pm$0.23 & 4.68$\pm$0.43  &  8.76$\pm$0.61 & 14.5$\pm$2.5 \\
 0.24 &	0.224$\pm$0.042	& 1.23$\pm$0.20 & 2.78$\pm$0.26 & 5.12$\pm$0.48  &  8.16$\pm$0.64 & 13.1$\pm$2.3 \\
 0.32 &	0.170$\pm$0.047	& 1.33$\pm$0.21 & 3.10$\pm$0.29 & 5.66$\pm$0.50  &  8.84$\pm$0.66 & 15.8$\pm$2.6 \\
 0.40 &	0.201$\pm$0.051	& 1.57$\pm$0.23 & 3.27$\pm$0.29 & 5.44$\pm$0.53  & 10.03$\pm$0.71 & 16.2$\pm$2.9 \\
 0.48 &	0.236$\pm$0.052	& 1.72$\pm$0.25 & 3.65$\pm$0.32 & 6.10$\pm$0.58  & 10.55$\pm$0.75 & 20.2$\pm$2.8 \\
 0.56 &	0.197$\pm$0.058	& 1.66$\pm$0.25 & 3.95$\pm$0.35 & 7.56$\pm$0.63  & 12.76$\pm$0.91 & 18.1$\pm$3.1 \\
 0.64 &	0.231$\pm$0.060	& 1.85$\pm$0.24 & 4.24$\pm$0.34 & 8.27$\pm$0.74  & 14.88$\pm$1.12 & 24.9$\pm$3.1 \\
 0.72 &	0.244$\pm$0.061	& 2.03$\pm$0.28 & 4.27$\pm$0.38 & 9.74$\pm$0.88  & 16.58$\pm$1.18 & 21.6$\pm$3.2 \\
 0.80 &	0.259$\pm$0.061	& 2.19$\pm$0.29 & 5.36$\pm$0.41 & 10.45$\pm$0.88 & 17.26$\pm$1.22 & 25.5$\pm$3.4 \\
 0.88 &	0.245$\pm$0.063	& 1.69$\pm$0.27 & 5.22$\pm$0.43 & 11.10$\pm$0.92 & 17.52$\pm$1.28 & 27.5$\pm$3.3 \\
 0.96 &	0.223$\pm$0.064	& 2.40$\pm$0.30 & 5.61$\pm$0.46 & 11.76$\pm$0.95 & 19.98$\pm$1.41 & 26.8$\pm$3.5 \\
\hline

\end{tabular}
\end{center}
\end{table}

\subsection{Dalitz plots and invariant mass distributions}

Dalitz plots and $M_{pp}$ invariant mass distributions of acceptance corrected 
and kinematically fitted $pp\pim$ events are presented in Figs.~8 and 9 together 
with statistical errors in the latter figure. In each case four Q bins 2\,MeV wide 
were chosen, the center values are indicated in each frame. The kinematical 
limits of the Dalitz plots given by the solid lines were calculated for these 
values; due to the rapidly growing phase space some data extend over these 
border lines. Here the size of the squares is a measure of the count rate. Each 
plot is almost uniformly covered with the exception of the area in the upper 
left corner where strong FSI effects between the reaction protons were expected. 
Also some lowering in yield is observed in the opposite corner which we attribute
to the asymmetries found in the pion angular distributions. No enhancements due 
to the $\Delta$ resonance are visible.

\begin{figure}[!htb]
\vspace*{-0.2cm}
\begin{center}
\epsfig{file=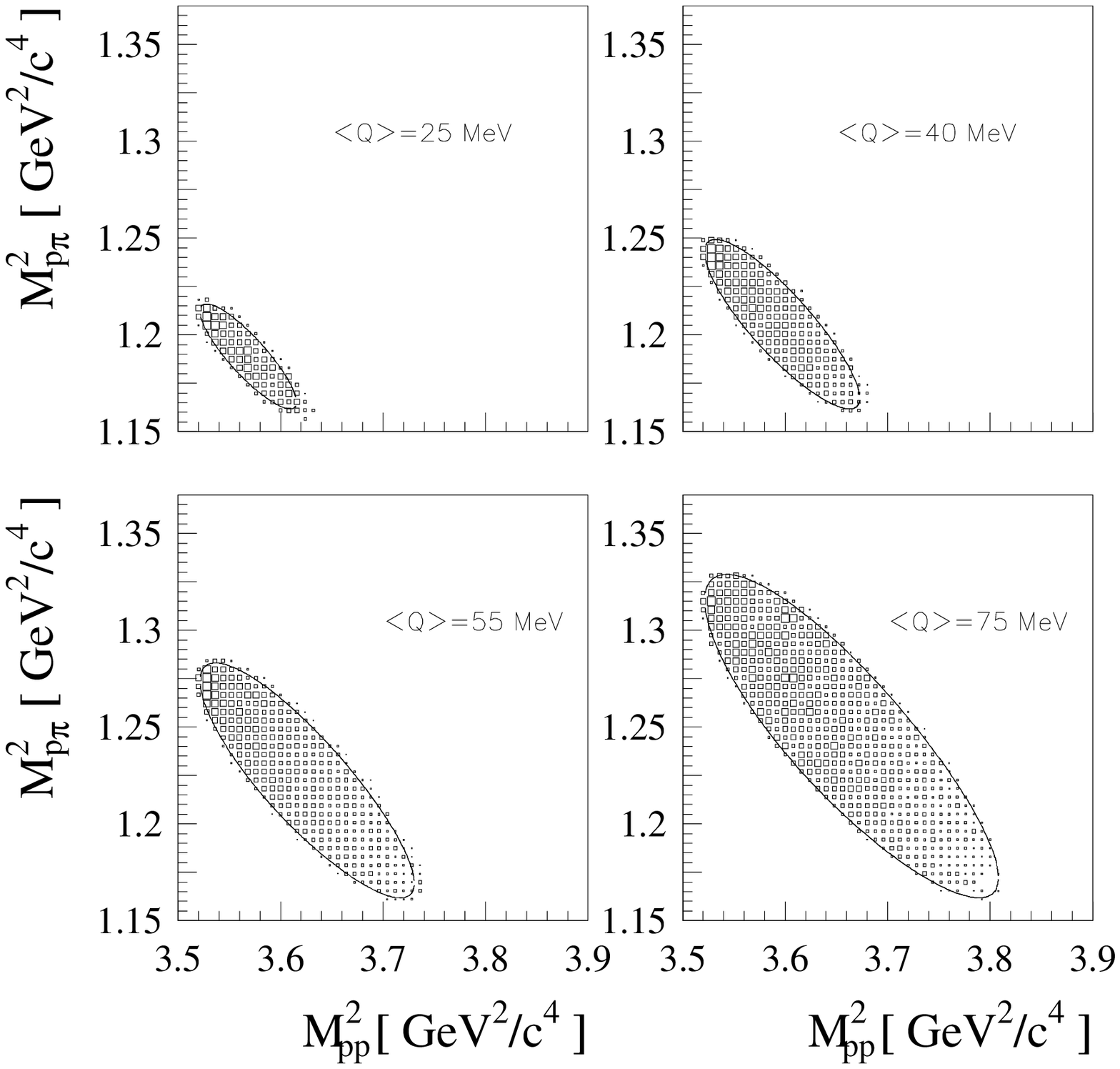,scale=0.6}
\end{center}
\vspace*{-0.8cm}
\caption{\it Experimentally deduced Dalitz plots for the quasifree reaction 
$np\to pp\pim$ at four 2\,MeV wide Q bins. The solid lines denote the 
kinematical limits.}

\end{figure}

\begin{figure}[!htb]
\vspace*{-0.2cm}
\begin{center}
\epsfig{file=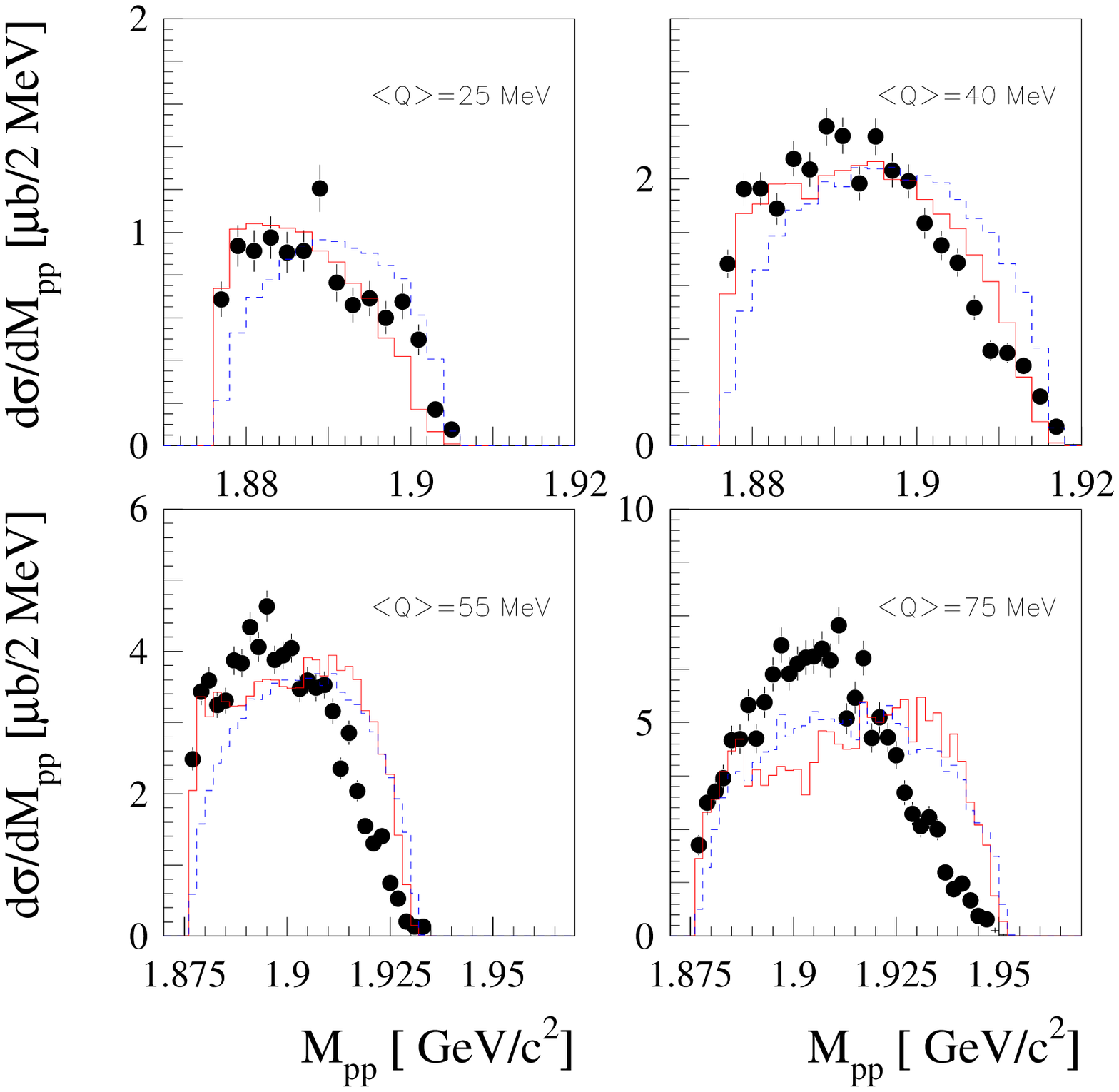,scale=0.6}
\end{center}
\vspace*{-0.8cm}
\caption{\it (Color online) Proton-proton invariant mass distributions obtained 
at four Q bins together with data from our Monte Carlo simulation. The dashed 
(blue) lines denote the results as found for phase space distributed events, the 
solid (red) curves give the ones where FSI effects with standard values for 
scattering length and effective range have additionally been included (see text).} 

\end{figure}

\newpage

The $d\sigma/dM_{pp}$ invariant mass distributions shown in Fig.~9 were plotted 
on a linear $M_{pp}$ scale, also given are the results obtained from our Monte
Carlo simulation (solid and dashed lines). In all cases large deviations are 
observed between experimental and simulated data, as long as purely phase
space distributed events were considered (dashed lines). Incorporating FSI 
effects into our MC simulations by using the formalism of Watson \cite{wats} 
and Migdal \cite{migd} which was later refined by Morton \cite{mort} the 
distributions given by the solid lines were found. We calculated additional 
weight factors $w_{fsi}$ given in a simplified form as

\vspace{-8mm}
\begin{center}
\begin{equation}
w_{fsi} = 1 + f_{pp} \cdot C^2 \cdot
\lbrack C^4 \cdot T^{CM}_{pp} + \frac{(\hbar c)^2}{m_p c^2} \left(
\frac{m_p c^2} {2(\hbar c)^2}r_0 \cdot T^{CM}_{pp} -\frac{1}{a_0}
\right) ^2 \rbrack ^{-1},
\end{equation}
\end{center}

where $T^{CM}_{pp}$ denotes the $pp$ center of mass kinetic energy $T^{CM}_{pp} 
= (M_{pp} -2m_p)\,c^2$ and $C^2$ the Coulomb
penetration factor

\vspace{-10mm}

\begin{center}
\begin{equation}
C^2 = \frac{2\pi \cdot \gamma_p}{e^{2\pi \gamma_p} - 1}
\end{equation}
\end{center}

\vspace{-4mm}

with $\gamma_p = \frac{\alpha \cdot \mu_{pp} \cdot c}{p_{pp}}$. Here $\alpha$ 
is the fine structure constant, $p_{pp} = \sqrt{2 \mu_{pp} T^{CM}_{pp}}$ and 
$\mu_{pp}$ is the reduced mass of the $pp$-system. The strength factor $f_{pp}$ 
is a measure of the contributing $Ss$ and $Sp$ partial waves and is adjusted for 
each $Q$ interval. From literature we took the standard values $a_0$=-7.83 fm
and $r_0$=2.8 fm \cite{fsi} as input parameters for the scattering length and 
effective range, respectively, for the two protons in the $^1S_0$ state. The 
agreement for the two lowest $Q$ bins, where the relative weight of the 
``diproton'' $1S_0$-state is high, is very good. In case of the two higher bins 
this simple ansatz, however, only succeeds in reproducing the rise at 
$M_{pp}\,=\,2m_p$.

\subsection{The isoscalar cross section}

Using equation (1) the isoscalar cross section $\sigma_{01}$ can be obtained 
from the measured cross sections for the $pp\to pp\piz$ and the $np\to pp\pim$
reactions. The isospin I\,=\,0 partial waves of type $Sp$ ($^3S_1\to$$^1S_0p_1$
and $^3D_1\to$$^1S_0p_1$) which are forbidden in the $pp\to pp\piz$ reaction due 
to the Pauli principle are generally believed \mbox{$\lbrack$2,6,10,11$\rbrack$} 
to dominate $\sigma_{01}$ in the threshold region. In the $np$ reaction they 
interfere with the isospin I\,=\,1 wave $^3P_0\to$$^1S_0s_0$  and thus are
responsible for the strong asymmetries in the $\pim$ angular distributions at 
very low excess energies. In the literature one finds a vast amount of data for 
both reactions and, as has been shown \cite{daum1}, the extracted $\sigma_{01}$ 
shows the expected $\eta^4$ dependence \cite{rose}, at least for $\eta >0.5$ 
($Q\,>17$\,MeV). We contend that the deviations quoted for smaller $\eta$ 
values are the result of wrong cross section data. As can be seen from (1), in 
order to obtain a finite $\sigma_{01}$ the $np$ cross section must at least be 
half as large as the one for the $pp$ reaction. For $\eta$\,=\,0.34 $\sigma_{np}$
is given as 1.43\,$\mu b$ \cite{daum1}. Recently our group at COSY reported new 
data for the $pp\to pp\piz$ reaction which exceeded the published ones from IUCF 
\cite{iucf} and CELSIUS \cite{cels} by roughly 50$\%$, the discrepancy could be 
shown to originate from an underestimation of the $pp$ final-state interaction 
\cite{cosy}. We found a cross section $\sigma_{pp}$\,=\,3.72$\pm 0.3 \mu b$ at 
$\eta$\,=\,0.35 which is 2.6 times larger than $\sigma_{np}$ and as such would 
leave no room for $\sigma_{01}$. This, however, is in contradiction to the 
asymmetries observed for the $\pim$ angular distributions, which are only 
possible with interfering $Ss$ and $Sp$ partial waves. In addition to assuming a 
wrong cross section measurement one should also consider a wrong beam energy 
determination. The uncertainty in neutron energy at the NA2 beam facility at PSI 
is given as 3 MeV ($\sigma$) for $E_n$\,=\,287\,MeV \cite{daum1}. An error in 
quoted beam energy of this size could possibly explain the observed deviations 
in the very close to threshold region where the cross section of almost any 
reaction rises dramatically.

\subsection{Summary}

With a deuteron beam at 1.85\,GeV/c impinging on a liquid hydrogen target the 
quasi-free $np\to pp\pim$ reaction was studied for excess energies $Q$ up to 90 
MeV. The data were analyzed in the framework of the spectator model where the 
proton is assumed to be an unaffected spectator staying on-shell throughout the 
reaction. Tagging the spectator proton in the forward scintillator hodoscope of 
our COSY-TOF spectrometer allowed to determine such parameters as effective mass 
and momentum of the off-shell neutron at the time of the reaction. We have 
measured angular distributions and invariant mass distributions of the reaction 
products and have set up Dalitz plots for several $Q$ bins distributed evenly 
over the whole excess energy range. The data were compared to results derived 
from Monte Carlo simulations and to data taken from the literature. In general 
good agreement was found, the large asymmetries observed previously for the 
$\pim$ angular distributions could be confirmed. Final-state interaction effects 
between the reaction protons were found even at the highest excess energies. 
The $d\sigma/dM_{pp}$ invariant mass distributions at 25 and 40\,MeV which are 
governed by the ``diproton'' $^1S_0$-state could be reproduced by the Monte 
Carlo simulations when incorporating FSI effects in the formalism of 
refs.~\mbox{$\lbrack$23-25$\rbrack$} with standard values for scattering length 
and effective range. Sizeable $d$-wave contributions were observed in the angular
distribution of the relative proton momentum at $Q$\,=\,82\,MeV. In view of new 
cross section data of the $pp\to pp\piz$ reaction, reported deviations of the 
isoscalar cross section $\sigma_{01}$ from an $\eta^4$ dependence \cite{daum1} 
were explained as stemming most probably from small errors in beam energy.

%\vspace*{1.0cm}

\subsection*{Acknowledgements}

The big efforts of the COSY crew in delivering a low-emittance deuteron beam
is gratefully acknowledged. Helpful discussions with C.~Hanhart, H.~Lacker, 
P.~Moskal and C.~Wilkin are very much appreciated. Special thanks are due to 
G.~Sterzenbach who as head of the workstation group provided continuous help in 
case of problems with the system. Financial support was granted by the German 
BMBF and by the FFE fund of the Forschungszentrum J\"ulich.

\end{sloppypar}

\end{document}